\pdfoutput=1
\documentclass[a4paper,fleqn,usenatbib]{mnras}

\usepackage{amsmath,txfonts}

\usepackage[T1]{fontenc}
\usepackage{ae,aecompl}

\usepackage{graphicx}	
\usepackage{amsmath}	
\usepackage{siunitx}	

\renewcommand{\vec}[1]{\mathbfit{#1}}
\renewcommand{\exp}[1]{\text{e}^{#1}}
\newcommand{\I}{\text{i}}
\newcommand{\diffd}{\text{d}}
\newcommand{\mathlabel}[1]{\text{#1}}
\let\oldcdot\cdot
\renewcommand{\cdot}{\boldsymbol{\oldcdot}}

\newcommand{\vabs}[1]{\lVert #1\rVert}
\newcommand{\abs}[1]{\lvert #1\rvert}

\newcommand{\vN}[1]{\sqrt{1-\vabs{#1}^2}}
\newcommand{\vn}{\vN{\vec{l}}}
\newcommand{\gt}{\widetilde{g}}
\newcommand{\Ft}{\widetilde{F}}
\newcommand{\Dl}{\Delta\vec{l}}
\newcommand{\ldd}{\vec{l}\cdot\vec{d}}
\newcommand{\tfov}{\theta_{\text{fov}}}
\newcommand{\tres}{\theta_{\text{res}}}
\newcommand{\Qfov}{Q_{\text{fov}}}

\let\eps\epsilon
\newcommand{\epsl}{\eps(\vec{l})}

\title{Approximating W projection as a separable kernel}

\author[Bruce Merry]{
Bruce Merry$^{1}$\thanks{E-mail: bmerry@ska.ac.za}
\\
$^{1}$SKA South Africa, 3rd Floor, The Park, Park Road, Pinelands, 7405, South Africa
}

\date{Accepted 2015 November 20. Received 2015 November 20; in original form
2015 September 10}

\pubyear{2015}

\begin{document}
\label{firstpage}
\pagerange{\pageref{firstpage}--\pageref{lastpage}}
\maketitle

\begin{abstract}
W projection is a commonly-used approach to allow interferometric imaging to
be accelerated by Fast Fourier Transforms (FFTs), but it can require a huge amount of storage for
convolution kernels. The kernels are not separable, but we show that they can
be closely approximated by separable kernels. The error scales with the fourth
power of the field of view, and so is small enough to be ignored at mid to
high frequencies.
We also show that hybrid imaging algorithms combining W projection with
either faceting, snapshotting, or W stacking allow the error to be made
arbitrarily small, making the approximation suitable even for
high-resolution wide-field instruments.
\end{abstract}




\section{Introduction}\label{sec:intro}
In interferometric imaging, the relationship between sampled visibilities and
the image plane is \emph{almost} a Fourier transform, but with a
term that depends on $w$ -- the dot product of the baseline vector with the
unit vector towards the phase centre. A number of approaches have been
developed to deal with this troublesome $w$ term.

One of these is W projection
\citep{wprojection}, which converts the image-space multiplication by a phase
screen into a convolution with its Fourier transform in $uvw$ space.
This convolution is usually combined with an anti-aliasing filter, yielding a
combined kernel, the Gridding Convolution Function (GCF), for which
closed-form formulae are not known. Instead, the GCF is
sampled in three dimensions and stored in a lookup table. Depending on the
desired accuracy, these kernels can become excessively large. Apart from
requiring large amounts of memory or disk space to store, this can
also reduce performance as the samples are moved in and out of caches.

Our contribution is to recognize that the phase screen for a particular value
of $w$, and hence its Fourier transform, is very close to being separable,
i.e., an outer product of two one-dimensional functions. By approximating the
phase screen in this way, we can drastically reduce the storage required for
lookup tables. Conversely, it allows for much finer sampling of the kernel in
the same amount of memory, potentially improving accuracy.

Sec.~\ref{sec:background} recaps the basics of W projection, and introduces
the notation we use. In Sec.~\ref{sec:derivation} we
describe our separable approximation, and provide a theoretical analysis of
the error. The error in position scales with the fourth power of the field of
view, and so is acceptable for small to medium fields of view.

W projection can be combined with other imaging techniques, which we
discuss in Sec.~\ref{sec:other}. We show that this allows the error to be
reduced further, allowing our approach to be used even for wide fields of
view. Sec.~\ref{sec:computation}
discusses the effect on computation cost, and our conclusions are presented
in Sec.~\ref{sec:conclusions}.

\section{Background and notation}\label{sec:background}
We will focus mainly on transforming from sampled visibilities to a dirty
image. Nevertheless, the same techniques and analysis apply to prediction of
visibilities from a model image.

As usual, $u, v, w$ are baseline coordinates in a fixed coordinate system with
$w$ in the direction of the phase centre. We measure
$u, v, w$ in wavelengths rather than units of
distance. The corresponding direction cosines are denoted $l, m, n$, with
$n=\sqrt{1-l^2-m^2}$. For compactness of notation, we will also use
$\vec{l} = \begin{pmatrix}l & m\end{pmatrix}^T$ and
$\vec{u} = \begin{pmatrix}u & v\end{pmatrix}^T$ interchangeably.

We will ignore direction-dependent effects, and assume a
time- and baseline-independent perceived brightness distribution $I(\vec{l})$.
Let the $i$th visibility have coordinates $(u_i, v_i, w_i)$, visibility value
$V_i$ and weight $W_i$. The dirty image $I^{\mathlabel{D}}$ is given by
\begin{equation}
  \frac{I^\mathlabel{D}(\vec{l})}{n} = \sum_i W_iV_i \exp{2\pi \I(\vec{u}_i\cdot\vec{l} + w_i(n-1))}.
  \label{eq:dirty}
\end{equation}
The corresponding prediction of visibilities from a model $I$ is
\begin{equation}
  V_i = \iint \frac{I(\vec{l})}{n} \exp{-2\pi \I(\vec{u}_i\cdot\vec{l} +
  w_i(n-1))} \diffd\vec{l}.
\end{equation}

Let $g_w(\vec{l}) = \exp{2\pi \I w(n-1)}$ and $G_w(\vec{u})$ be its Fourier
transform, and let $V^{\mathlabel{W}}_i(u, v) = \delta(u-u_i)\delta(v-v_i)W_iV_i$.
Then
\begin{equation}
\begin{aligned}
  I^{\mathlabel{D}}(l, m)/n
  &= \sum_i \left(\iint V^{\mathlabel{W}}_i(\vec{u})\,\diffd\vec{u}\right) \exp{2\pi \I(\vec{u}_i\cdot\vec{l} + w_i(n-1))}\\
  &= \sum_i \left(\iint V^{\mathlabel{W}}_i(\vec{u}) \exp{2\pi \I\vec{u}\cdot\vec{l}}\,\diffd\vec{u}\right)g_{w_i}(\vec{l})\\
  &= \sum_i \mathbfss{F}^{-1}\big[V^{\mathlabel{W}}_i\big](\vec{l}) g_{w_i}(\vec{l})\\
  &= \sum_i \mathbfss{F}^{-1}\big[V^{\mathlabel{W}}_i * G_{w_i}\big](\vec{l})\\
  &= \mathbfss{F}^{-1}\Big[\sum_i V^{\mathlabel{W}}_i * G_{w_i}\Big](\vec{l}).
  \label{eq:wprojection}
\end{aligned}
\end{equation}
This is typically combined with an antialiasing kernel $C$ (with inverse
Fourier transform $c$) to give
\begin{equation}
  I^{\mathlabel{D}}(l, m)/n = \mathbfss{F}^{-1}\Big[\sum_i V^{\mathlabel{W}}_i * G_{w_i} * C\Big](l, m) / c(l, m).
\end{equation}

The two convolution kernels $G_w$ and $C$ are combined into a gridding
convolution function (GCF) $F_w$ by multiplying their inverse Fourier
transforms $g_w$ and $c$, and taking the Fourier transform of the
result.

\begin{figure}
  \centering
  \includegraphics{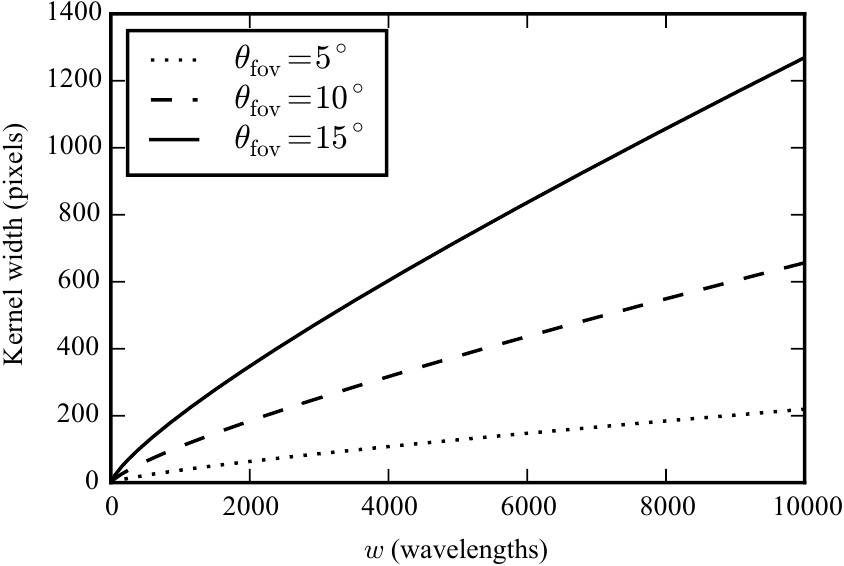}
  \caption{Support for the W kernel with truncation at $\eta = 0.01$ of the
   peak. $\tfov$ is the field of view.}
  \label{fig:wsize}
\end{figure}

Generating a kernel is expensive, so usually they are either fully
precomputed, or are generated on-demand then cached. The storage size is thus
an important consideration.  The antialiasing kernel $C$ is typically small
(e.g., $9\times 9$ pixels), so the support of the GCF is dominated by the
support of $G_w$. The support (in $uv$-plane pixels) necessary to represent the
function out to the fraction $\eta$ of the peak is \citep{wsupport}
\begin{equation}
  2\tfov\sqrt{\left(\frac{w\tfov}{2}\right)^2 + \frac{w^{3/2}\tfov}{2\pi\eta}}.
\end{equation}
Fig.~\ref{fig:wsize} shows typical values of this function for $\eta = 0.01$.

Kernels can be hundreds of pixels across, even for baselines of a few
kilometres. What is worse, they must be oversampled to avoid aliasing effects.
A typical oversampling factor is 8 \citep{romein-gridding}, which means that a
single kernel may require millions of samples and thus consume tens of
megabytes.

\citet{wsnapshots} suggest that the number of $w$ planes should be
\begin{equation}
  \frac{\pi w_{\text{max}}(\tfov)^2}{\sqrt{2\Delta A}}
\end{equation}
where $\Delta A$ is the tolerable loss of amplitude in the image plane due to
decorrelation effects. This can easily reach $10^4$ at low elevations with
$\Delta A = 0.01$, and storing all the kernels can require tens or even
hundreds of gigabytes. Our method can reduce memory usage by a
factor of 1000 or more, making precomputation practical.

\section{Derivation and analysis}\label{sec:derivation}
We aim to approximate $F_w$ as a separable function. We assume that the
antialiasing function $c$ is separable, and hence it suffices to approximate
$g_w$ as a separable function.

\begin{figure}
  \centering
  \includegraphics{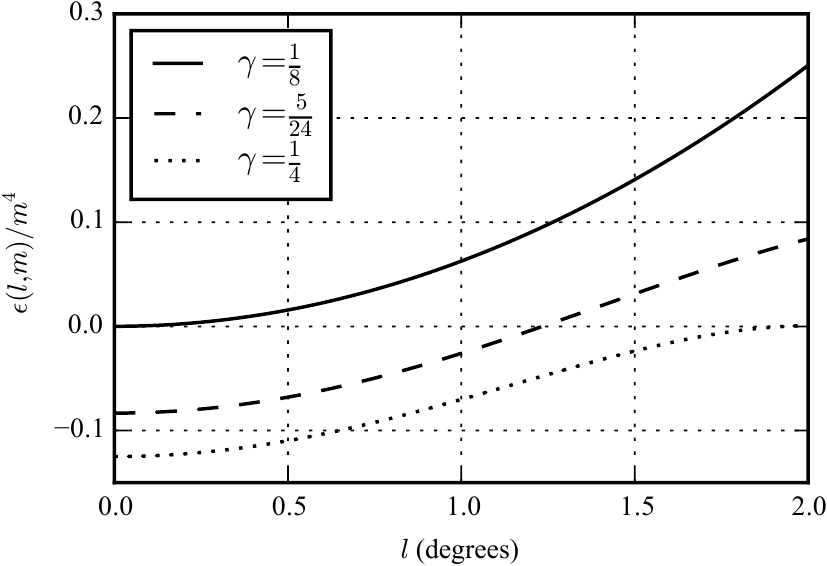}
  \caption{Phase error factor along the edge of an image with \ang{4} field of view
  ($m = \ang{2}$).}
  \label{fig:epsplot}
\end{figure}

Recall that $g_w(l, m) = \exp{2\pi \I w(n-1)}$ and $n=\sqrt{1-l^2-m^2}$.
Using the Taylor expansion
$\sqrt{1-x} = 1-\frac{1}{2}x - \frac{1}{8}x^2 + O(x^3)$, we get
\begin{equation}
  g_w(l, m) = \exp{2\pi \I w\big[-\tfrac12(l^2+m^2) -
  \tfrac18(l^2+m^2)^2+O(l^6+m^6)\big]}.
\end{equation}
The first few terms of the phase depend on either $l$ or $m$, but not
both, and these are the dominant terms when $\abs{l}, \abs{m} \ll 1$. This is
what makes $g$ approximately separable. Let
\begin{align}
  g^{\mathlabel{1}}_w(l) &= \exp{2\pi \I w\big[-\frac12 l^2 - \gamma l^4\big]}\\
  \gt_w(l, m) &= g^{\mathlabel{1}}_w(l)g^{\mathlabel{1}}_w(m),   \label{eq:gt}
\end{align}
where $\gamma$ is a tuning parameter we will discuss later. We approximate
$g_w$ by $\gt_w$. This approximation introduces an image-space phase error of
\begin{equation}
\begin{aligned}
  \Delta\phi(\vec{l}, w)
    &= 2\pi w\Big[
      \begin{aligned}[t]
        &-(\tfrac12 l^2+\gamma l^4+\tfrac12 m^2+\gamma m^4)\\
        &-(\sqrt{1-l^2-m^2}-1)\Big]
      \end{aligned}\\
    &= 2\pi w\Big[
      \begin{aligned}[t]
        &\tfrac12(l^2+m^2)+\tfrac18(l^2+m^2)^2+O(l^6+m^6)\\
        &-\tfrac12 l^2-\gamma l^4-\tfrac12 m^2-\gamma m^4\Big]
      \end{aligned}\\
    &= 2\pi w\Big[(\tfrac18-\gamma)(l^4+m^4) + \tfrac 14 l^2m^2 + O(l^6 + m^6)\Big].
  \end{aligned}
\end{equation}
Denote the factor inside the square brackets by $\epsl$. We would like to
minimize $\abs{\epsl}$ over the image. The largest values will clearly be along
the edges. Fig.~\ref{fig:epsplot} shows how $\epsl$ varies along an edge
for several values of $\gamma$. Ignoring the $O(l^6+m^6)$ term, the value
$\gamma=\frac{5}{24}$ minimizes the maximum error at
$\frac{1}{12}m_{\text{max}}^4$, while $\gamma=\frac{1}{8}$ favors accuracy on
the $l=0$ and $m=0$ axes and pushes the error into the corners.

We will now approximate the array as instantaneously co-planar (ignoring the
slight curvature due to the shape of the Earth), so that $w$ can be computed as
$w = \vec{d}\cdot\vec{u}$ for some time-varying $\vec{d}$. Note that
$\vabs{\vec{d}} = \cot a$ ($a$ being the elevation angle), so it is bounded by
the minimum elevation.
This allows us to recast the phase error as an instantaneous position error.
When substituting the approximation into \eqref{eq:wprojection}, we get
\begin{equation}
\begin{aligned}
  I^{\mathlabel{D}}(l, m)/n
  &= \sum_i W_iV_i \exp{2\pi \I\big[\vec{u}_i\cdot \vec{l} + w_i(\vn-1+\epsl)\big]}\\
  &= \sum_i W_iV_i \exp{2\pi \I\vec{u}_i\cdot\big[\vec{l} + \big(\vn-1+\epsl\big)\vec{d}\big]}.
\end{aligned}
\end{equation}
If we can find $\vec{l}'$ such that
\begin{equation}
  \vec{l}' + \big(\vN{\vec{l}'}-1\big)\vec{d} = \vec{l} + \big(\vn-1+\epsl\big)\vec{d}
  \label{eq:lprime}
\end{equation}
then the phase error is equivalent to shifting $\vec{l}'$ in the correct dirty
image to $\vec{l}$ in the approximated dirty image. In other words, at a
single point in time the approximation will cause a distortion in the image;
as the Earth rotates, $\vec{d}$ will change, causing sources to be smeared out.

If $\Dl = \vec{l}' - \vec{l}$ is sufficiently small, then this
error can reasonably be ignored. \citet{wsnapshots} show that
the relative loss in amplitude at $\vec{l}$ is proportional to
$\left(\frac{\vabs{\Dl}}{\tres}\right)^2$, assuming a parabolic
dirty beam, where $\tres$ is the resolution; it is
thus desirable that $\vabs{\Dl} \ll \tres$.
While \eqref{eq:lprime} can be solved
exactly (see Appendix~\ref{sec:proof}), we can obtain more insight from the
approximation
\begin{equation}
  \Dl \approx (1+\ldd/n)\epsl\vec{d} \approx \epsl\vec{d}
  \label{eq:deltal}
\end{equation}
which suggests that $\vabs{\epsl \vec{d}} \ll \tres$ is sufficient.

We can translate this into a relationship between parameters of the
instrument. The field of view $\tfov$ is given by $2\Qfov \frac{\lambda}{D}$,
where $\lambda$ is the wavelength, $D$ is the antenna/station diameter, and
$\Qfov$ is a constant that depends on the antenna (e.g., illumination
tapering) and the desired field of view relative to the primary beam size. For
example, $\Qfov = 1.22$ gives an image that encloses the first null of an
ideal Airy disk \citep[p.~41]{whitebook}. For resolution we use the
rule of thumb $\tres = \frac{3\lambda}{2B}$, where $B$ is the length of the
longest baseline; the actual constant factor depends on the $uv$ coverage and
imaging weights \citep[p.~131]{whitebook}. As noted above, $\vabs{\vec{d}} =
\cot a$. Thus,
\begin{equation}
\begin{aligned}
  \frac{\vabs{\Dl}}{\tres}
  &\approx \frac{\frac{1}{12}(\Qfov \frac{\lambda}{D})^4 \cot a}
          {\left(\frac{3\lambda}{2B}\right)}
  = \frac{1}{18}\Qfov^4 \lambda^3 D^{-4}B\cot a
\end{aligned}
\label{eq:errscale}
\end{equation}

To provide an example, let us consider some worst-case values for MeerKAT
(Karoo Array Telescope) \citep{meerkat-params,meerkat-fact-sheet}:
$\lambda = \SI{0.333}{\metre}$ (\SI{900}{\MHz}), $D = \SI{13.5}{m}$, $B = \SI{8}{\km}$,
$a = \ang{15}$ gives $\frac{\vabs{\Dl}}{\tres} \approx 0.0018 \Qfov^4$,
allowing for a large $\Qfov$ without problems.

Phase 1 of Square Kilometre Array mid-frequency instrument (SKA1-MID) will
support lower frequencies and much longer baselines \citep{ska1-params}: $D =
\SI{15}{\m}$, $B = \SI{150}{\km}$, $\lambda = \SI{0.857}{\m}$
(\SI{350}{\MHz}). In this case, \eqref{eq:errscale} gives $0.39 \Qfov^4$,
assuming the same $\ang{15}$ minimum elevation as for MeerKAT. This means that
the position error is of the same order as the beam size, which over time will
cause decoherence at the edges of the image.

Due to the cubic dependence on $\lambda$, our basic approach is applicable at
mid and high frequencies, but breaks down at lower frequencies. In general, we
see limited value to the basic approach below about $\SI{500}{\MHz}$, at least
for arrays of small dishes. However, combining W projection with other imaging
approaches allows the error to be made arbitrarily small, making our approach
suitable even at low frequencies. This is discussed further in
Sec.~\ref{sec:other}.

It is interesting to note that the analysis above applies to other phase
errors of the form $2\pi w\epsl$. In particular, if no correction for W
is done at all, then $\epsl = 1 - \vn$. Surprisingly, \eqref{eq:deltal} does
not depend on the array diameter or largest $w$ value, even though $w$ effects
are conventionally associated with long baselines \citep{wprojection}. Long
baselines play an indirect role, in that they give high resolution and hence
position errors are larger relative to the synthesized beam.

\section{Combination with other techniques}\label{sec:other}
W projection is only one approach to dealing with the $w$ term, and hence
allowing Fast Fourier Transforms to be used to accelerate imaging. Others
include faceting, snapshot imaging, and W stacking. In each case, it is
possible to hybridize the algorithm with W projection to combine the
computational strengths of both. W projection's strength is in handling small
$w$ values very accurately (limited only by sampling resolution), but it
becomes computationally very costly to handle large $w$ values as the
necessary support of the GCF grows. Hybrid techniques use alternative methods
to make coarse phase corrections, with the finer corrections left to W
projection.

Since the phase error in our method scales with $w$, it seems likely that
these hybrid methods would be particularly suitable since the $w$ values
involved would be reduced. The following subsections show that this is
indeed the case.

\subsection{Faceting}
Faceting has a number of advantages for imaging: each facet requires less
working memory, and the reduction in field of view allows $w$ effects to be
ignored or to be corrected more easily. Here we will discuss faceting in which
the field of view of each facet is small, but not so small that $w$ effects can
be completely ignored. This is an important domain, because there is a fixed
cost per facet to adjust and grid the visibilities, and hence a large number
of tiny facets can be computationally costly.

In classical faceting, the coordinate system is rotated and the phase of
visibilities is adjusted so that each facet has its own phase centre. In terms
of $w$ effects, each facet behaves as an independent image with a narrower field
of view. Since the error scales with the fourth power of field of view, using
even a small number of facets will greatly reduce the errors.

More recently, `image-plane' faceting \citep{aips113} applies a
non-orthogonal change of coordinate system so that the facets are all part of
a single plane tangent to the celestial sphere at the original phase centre,
rather than each being tangent at the facet centre. This simplifies generation
of a single image from the facets, but complicates our analysis.

\citet{aips113} derive the equations for image-plane faceting using only a
first-order Taylor approximation to the $w$ factor. In their derivation, all
$w$ effects are corrected by suitable adjustments to $\vec{u}$. For a hybrid
W-projection/faceting algorithm, we must compute a second-order approximation.

Let $h(\vec{x}) = \sqrt{1-\vabs{\vec{x}}^2}$, let $\vec{l}_0$ be the facet
centre, with $n_0=h(\vec{l}_0)$, and let $\vec{l}$ refer to the position relative to the
\emph{facet} centre, such that the position relative to the original phase
centre is $\vec{l}+\vec{l}_0$. The modified $w$ correction is
\begin{equation}\label{eq:phase-correction-facet}
  g_w(\vec{l}) = \exp{2\pi \I w(n-n_0)} = \exp{2\pi \I w[h(\vec{l}_0+\vec{l}) - h(\vec{l}_0)]}
\end{equation}
We now approximate $h$ using a Taylor polynomial about $\vec{l}_0$:
\begin{equation}
\begin{aligned}
  h(\vec{l}_0 + \vec{l}) - h(\vec{l}_0)
  &\approx \nabla h(\vec{l}_0)\vec{l}
  + \tfrac{1}{2}\vec{l}^T \mathbfss{H}h(\vec{l}_0)\vec{l}\\
  &= -n_0^{-1}\vec{l}_0^T\vec{l}
     + \tfrac12\vec{l}^T\big[-n_0^{-1}I - n_0^{-3}\vec{l}_0\vec{l}_0^T\big]\vec{l}\\
  &= \begin{aligned}[t]
        &-n_0^{-1}\vec{l}_0^T\vec{l}
        -\tfrac12(n_0^{-1}+l_0^2n_0^{-3})l^2\\
        &-\tfrac12(n_0^{-1}+m_0^2n_0^{-3})m^2
        -n_0^{-3}l_0m_0lm.
     \end{aligned}
\end{aligned}
\end{equation}
The initial term is absorbed into the coordinate transformation $\vec{u}' =
\vec{u} - n_0^{-1}\vec{l}_0w$. The next two terms can be handled by a
separable kernel, while the final term is not separable. We replace
\eqref{eq:gt} by
\begin{align}
  g^{\mathlabel{l}}_w(l) &= \exp{2\pi \I w\big[-\tfrac{1}{2}(n_0^{-1}+l_0^2n_0^{-3})l^2\big]}\\
  g^{\mathlabel{m}}_w(m) &= \exp{2\pi \I w\big[-\tfrac{1}{2}(n_0^{-1}+m_0^2n_0^{-3})m^2\big]}\\
  \gt_w(l, m) &= g^{\mathlabel{l}}_w(l)g^{\mathlabel{m}}_w(m).\label{eq:gt-facet}
\end{align}
There are a few noteworthy differences from the case of classical faceting.
First, the component functions are now facet-dependent, and different for
$l$ and $m$. We have also used a lower-order approximation, because without
being able to incorporate the $-n_0^{-3}l_0m_0lm$ term, there is little point
in trying to compute higher-order terms. This also means that for a given
total field of view, the error now only scales with the square rather than the
fourth power of the facet width.

\begin{figure*}
  \centering
  \includegraphics{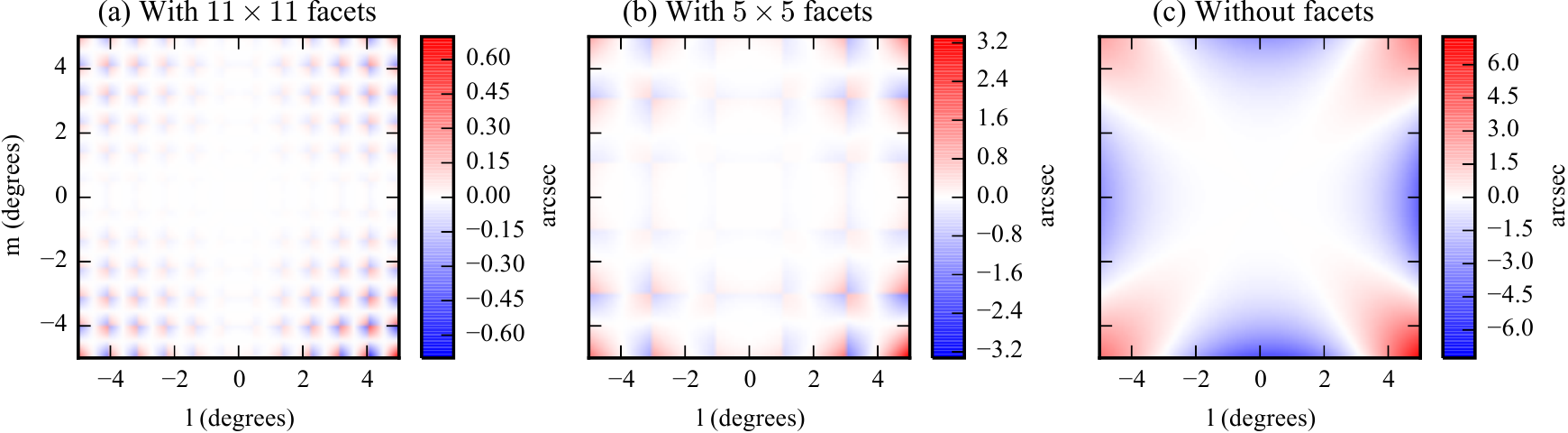}%
  \caption{Position error when using \eqref{eq:gt-facet} with facets (a and
    b) versus \eqref{eq:gt} and no facets (c).}%
  \label{fig:facet-error}
\end{figure*}

Fig.~\ref{fig:facet-error} shows the position error at \ang{15} elevation and
a \ang{10} field of view,
using $11\times 11$ and $5\times 5$ grids of
facets. In the $11\times 11$ case, the error is reduced by an order of
magnitude compared to the unfaceted approach. However, the discontinuities at
the boundaries between facets may be problematic, since a source on a boundary
will appear in different positions in the two facets.

Our claim that the error scales quadratically with facet size holds between
\ref{fig:facet-error}a and \ref{fig:facet-error}b, but \ref{fig:facet-error}c
does not follow the same scaling. This is for two reasons. Firstly,
\ref{fig:facet-error}c uses \eqref{eq:gt} rather than \eqref{eq:gt-facet}, and
the former partially corrects for fourth-order errors. Secondly, the claim
only holds for facets with the same centre; since \ref{fig:facet-error}c has
the facet centre at the phase centre, the errors are smaller.

\subsection{W snapshots}
Snapshot imaging corrects the position error introduced by $w$ terms in image
space. Since the position error is time-varying, an image is created for a
small window of time, then reprojected to correct the distortions. Over a longer
observation, many images are made and added together. For each snapshot, a
single $\vec{d}_0$ value is found such that $w \approx \vec{u}\cdot\vec{d}_0$,
and then after creating the image, the image is reprojected so that
$\vec{l}^{\mathlabel{s}} = \vec{l} + \vec{d}_0(n-1)$ is mapped to $\vec{l}$.

W snapshots \citep{wsnapshots} is a hybrid method combining snapshot imaging
with W projection. The snapshot warping corrects for an average position error
over the snapshot duration, while W projection fixes up residual errors caused
by long snapshots or by non-planar arrays. This allows the snapshot duration
to be much longer than for traditional snapshot imaging, while still reducing
the gridding costs by keeping $w$ values small.

The component $2\pi \I\vec{u}\cdot\vec{d}_0(n-1)$ of the phase
is absorbed into the reprojection, leaving a residual of $2\pi
\I\vec{u}\cdot(\vec{d}-\vec{d}_0)(n-1)$ to be handled by W
projection\footnote{For analysis we will assume a coplanar array, but
deviations from planarity will also appear in the residual.}. It
follows that the remaining position error from using a separable kernel is
\begin{equation}
  \Dl \approx (1+\vec{l}\cdot(\vec{d}-\vec{d}_0))\epsl(\vec{d}-\vec{d}_0).
\end{equation}

There is another source of phase error, not discussed by \citet{wsnapshots}:
W projection does not account for the reprojection\footnote{This could be
  fixed by generating a custom kernel for each value of $\vec{d}_0$. We have
not analyzed this approach.}, and hence the actual phase
correction applied is $2\pi \I\vec{u}\cdot(\vec{d}-\vec{d}_0)(\sqrt{1-\vabs{\vec{l}^{\mathlabel{s}}}^2}-1)$.
This will introduce a phase error of
\begin{equation}
\begin{aligned}
  &\phantom{{}\approx{}} 2\pi\vec{u}\cdot(\vec{d}-\vec{d}_0)\left(\sqrt{1-\vabs{\vec{l}^{\mathlabel{s}}}^2}-\sqrt{1-\vabs{\vec{l}}^2}\right)\\
  &\approx 2\pi\vec{u}\cdot(\vec{d}-\vec{d}_0)(-\vec{l}\cdot\vec{d}_0(n-1))\\
  &\approx 2\pi\vec{u}\cdot(\vec{d}-\vec{d}_0)(\tfrac12\vabs{\vec{l}}^2\vec{l}\cdot\vec{d}_0)
\end{aligned}
\end{equation}
For comparison, the phase error associated with separation of the kernel, when
$m=0$ and $\gamma = \frac{5}{24}$, is approximately
$2\pi \vec{u}\cdot(\vec{d}-\vec{d}_0)\epsl \approx
2\pi \vec{u}\cdot(\vec{d}-\vec{d}_0)(\frac{1}{12}\vabs{\vec{l}}^4)$. Unless
the phase centre is close to the zenith, $\vabs{\vec{d}_0} \gg \vabs{\vec{l}}$, and so the
error due to separation will be insignificant compared to this other error. Of
course, both errors can be made arbitrarily small by limiting the snapshot
length, although in the limit the assumption of a coplanar array
will break down.

\subsection{W stacking}
W stacking \citep{wsclean} is similar to snapshot imaging in that visibilities
are partitioned, and within each partition the average $w$ effects are corrected in
image space. Instead of partitioning visibilities by time, they are
partitioned by $w$, and the image-space adjustment is a phase correction
rather than a reprojection. As with snapshotting, it is possible to use
coarser partitioning, and hence fewer Fourier transforms, by using W
projection to deal with residual $w$ terms.

Unlike in the previous cases, the $w$ projection term is no longer directly
correlated with $\vec{u}$ at a point in time. The errors will thus cause
smearing rather than a systematic shift in position. Let $\Delta w$ be the
residual $w$ that will be adjusted by $w$ projection. If the slices are evenly
spaced $2w_\text{max}$ apart, then $\abs{\Delta w} \le w_\text{max}$. With
sufficiently many slices, $\Delta w$ can be treated as a uniformly-distributed
random variable.

A phase error of $\phi$ will cause a relative amplitude loss of
$1-\cos\phi$ at the position of a point source. Hence,
the expected loss of amplitude is
\begin{equation}
\begin{aligned}
  \big\langle 1-\cos 2\pi \Delta w\epsl\big\rangle
  &\approx \big\langle \tfrac{1}{2}(2\pi \Delta w\epsl)^2\big\rangle\\
  &= 2\pi^2 \epsl^2\big\langle \Delta w^2\big\rangle\\
  &= \tfrac{2}{3}\pi^2 \epsl^2 w_\text{max}^2.
\end{aligned}
\label{eq:amplitude-loss}
\end{equation}

Let us re-evaluate the SKA1-MID example from Sec.~\ref{sec:derivation}, with a
\ang{10} field of view (corresponding very roughly to the second null of the
primary beam).
In this case, $\abs{\epsl} < 10^{-5}$ everywhere in the image, so provided that
$w_\text{max} \ll 10^5$, the loss in amplitude will be minimal. At
\SI{500}{MHz}, the longest baseline is $2.5\times 10^5$
wavelengths, so this will not require an unreasonable number of slices.

A potential issue with a core-heavy array is that if the $w=0$ slice is too
wide, it will contain a substantial fraction of the visibilities. These
visibilities contribute to the dirty image exactly as if no W stacking was
applied, and hence create a ghost source at $\vec{l}'$ as in
equation~\eqref{eq:lprime}. This suggests placing a large number of slices
close to $w=0$, with wider-spaced slices at larger $w$ values. Adapting slice
width to the visibility density will also reduce computation time, because
narrow slices need less support for the convolution kernel, hence speeding up
gridding, but sparse regions benefit from fewer, wider slices to reduce
Fourier transform costs.


\section{Computation cost}\label{sec:computation}
With our approach, the gridding convolution function becomes a separable function
$\Ft_w(u, v) = F^{\mathlabel{1}}_w(u)F^{\mathlabel{1}}_w(v)$.
Updating the grid cell at $(u, v)$ with visibility
$i$ now requires computing $V_i F^{\mathlabel{1}}_{w_i}(u-u_i) F^{\mathlabel{1}}_{w_i}(v-v_i)$
rather than $V_i F_{w_i}(u-u_i, v-v_i)$. This would
appear to require double the number of multiplications. However, the partial
product $V_i F^{\mathlabel{1}}_{w_i}(u-u_i)$ depends only on $u$ and not $v$, and hence can
be computed once and reused for every value of $v$ in the support of the
kernel. The number of multiplications thus only increases by a factor of
$1+\frac{1}{\abs{F^{\mathlabel{1}}_w}}$, where $\abs{F^{\mathlabel{1}}_w}$ is the width of the kernel in grid cells.

It is also important to emphasize that floating-point operations are not
necessarily the bottleneck in gridding: \citet{romein-gridding} and
\citet{muscat-gridding} both report that performance is limited by the memory
system in their respective GPU-accelerated gridders. Muscat reports
that loading the kernel data from texture memory is the limiting factor, and
our smaller kernel is likely to improve cache efficiency here. Our
prototype GPU gridder using a separable kernel achieves greater efficiency
than the published figures for either of these implementations, but
further work is required to establish whether this is due to other factors.

\section{Conclusions and future work}\label{sec:conclusions}

We have demonstrated a method to approximate the GCF for W projection which
reduces the lookup table from three to two dimensions, leading to a massive
reduction in memory. This makes it possible to store more $w$ planes, to make
better usage of caches, and possibly to keep the entire kernel in a smaller
but faster level of a memory hierarchy. We have focused on dirty imaging, but
the same approximation can be used for prediction, allowing for Cotton--Schwab
major cycles \citep{cotton-schwab}.

The approximation introduces a phase error, which is linear in $w$ and hence
translates to an instantaneous position error for a coplanar array. The error is small,
scaling with the fourth power of the field of view. The approach is thus
practical for mid- and high-frequency dish arrays, even with pure W
projection, provided that baselines are no more than a few tens of kilometres
long. The error can also be made arbitrarily
small by using faceting, snapshotting or W stacking
with a sufficient number of facets, snapshots or stack
slices. This allows the basic idea to be extended to higher resolutions and
lower frequencies, such as is envisioned for SKA1-MID. It remains future work
to investigate whether wide-field, low-frequency aperture arrays can be
supported without requiring an excessively large number of
facets/snapshots/slices that would harm the overall performance.

While we believe that our approach will increase performance due to reduced
pressure on memory systems, we have not yet implemented both versions of the
kernel in a single gridding application to provide a fair comparison.

An idea that merits further investigation is to obtain better accuracy by
storing and applying the correction $F_w - \Ft_w$. This will potentially
require far less support and/or sampling rate than the original GCF, and hence
can be stored in a small yet full-dimension table. It may also be possible to
approximate this correction as another separable function, but rotated
$\ang{45}$ to the original.

Another method used in interferometric imaging is A projection, in which
primary beam correction is done in the $uv$ plane by convolution
\citep{aprojection}. This may also be combined with W projection, and one may
naturally ask whether our technique can be applied to the combination.
Unfortunately this is unlikely to be as successful, as the primary beam
rotates on the sky and is unlikely to be separable in all orientations. Our
technique can still be used with other approaches to primary beam correction,
such as image-space correction for each snapshot with W snapshots.




\bibliographystyle{mnras}
\bibliography{separable-w.bib}



\appendix

\protected\def\Dlmath{$\Dl$}
\section{Proof of approximation for \Dlmath}\label{sec:proof}
Here we will derive an exact solution to \eqref{eq:lprime} and justify the
approximation given in \eqref{eq:deltal}. Rearranging \eqref{eq:lprime} gives
\begin{equation}\label{eq:dl}
  \Dl = \left[\vn-\vN{\vec{l}+\Dl}+\epsl\right]\vec{d}.
\end{equation}
It is immediately clear that $\Dl$ is a multiple of $\vec{d}$, so let $\Dl =
y\vec{d}$. Let
\begin{equation}
k = \vn + \epsl = n + \epsl.
\end{equation}
Then
\begin{align}
&&         y &= \vn - \vN{\vec{l}+y\vec{d}}+\epsl\\
\iff&& y - k &= -\vN{\vec{l}+y\vec{d}} \label{eq:ymk}\\
\implies&& (y-k)^2 &= 1-\vabs{\vec{d}}^2y^2 - 2(\ldd)y - \vabs{\vec{l}}^2\\
\iff&&     0 &=
  \begin{aligned}[t]
    &(1 + \vabs{\vec{d}}^2)y^2 + 2(\ldd-k)y\\
    &+(k^2+\vabs{\vec{l}}^2-1)
  \end{aligned}\\
\iff&&     0 &= 
  \begin{aligned}[t]
    &(1 + \vabs{\vec{d}}^2)y^2 + 2(\ldd-k)y\\
    &+\big(2n + \eps(\vec{l})\big)\eps(\vec{l}).
  \end{aligned}
\end{align}
This is a quadratic equation in $y$, which can easily be solved. It is
possible that both two roots satisfy \eqref{eq:ymk}, but we are only
interested in the one with the smaller absolute value.

We now turn to approximating $y$. To get an initial estimate, we ignore the
term $\vn-\vN{\vec{l}+\Dl}$, giving $y_0 = \epsl.$ To improve this, we note
that
\begin{equation}
\begin{aligned}
  \vN{\vec{l}+\Dl}
  &= \vN{\vec{l}+y\vec{d}}\\
  &= \sqrt{1-\big(\vabs{\vec{d}}^2y^2 + 2(\ldd)y + \vabs{\vec{l}}^2\big)}\\
  &\approx \vn - \left(\ldd/\vn\right)y\\
  &= n - \big(\ldd/n\big)y
\end{aligned}
\end{equation}
where the approximation is a linear Taylor polynomial around $y=0$.
Substituting this back into \eqref{eq:dl} gives
\begin{equation}
  y \approx \big(\ldd/n\big)y + \epsl
\end{equation}
and if we use $y_0 = \epsl$ in place of $y$ in the right-hand side, we get the
improved approximation
\begin{equation}
  y_1 = (1 + \ldd/n)\epsl.
\end{equation}


\bsp	
\label{lastpage}
\end{document}